# Paradoxical Coffee-Stain Effect Driven by the Marangoni Flow Observed on Oil-Infused Surfaces[1]

Gilad Chaniel[a], Mark Frenkel[b], Victor Multanen[b], Edward Bormashenko[b*]

[a]Ariel University, Natural Science Faculty, Physics Department, 407000, P.O.B. 3, Ariel, Israel

[b]Ariel University, Engineering Faculty, Chemical Engineering and Biotechnology Department, 407000, P.O.B. 3, Ariel, Israel

E-mail: edward@ariel.ac.il

**ABSTRACT**

Formation of coffee stain deposits under evaporation of droplets containing aqueous solution of salts placed on silicone-oil impregnated substrates was observed. The formation of ring-like deposits was registered for various molar concentrations of salts for the droplets of 5–300 µl in volume. The effect occurred when the contact line was de-pinned, and the evaporation from the edge of a droplet was stopped by the silicone oil. The formation of the coffee stain deposit is related to the soluto-capillary Marangoni flow; the influence of the thermo-capillary flow taking place in parallel is negligible.

**Keywords**: coffee-stain deposit; Marangoni flow; oil-impregnated substrate; evaporation; contact line.

1. Introduction

Evaporating drops of colloidal suspensions and solutions of non-volatile species leave behind ring-like solid residues along the contact line. This is the coffee stain effect, named after the most widely known representative of this class of structures. In spite of its apparent simplicity the understanding of the coffee-stain calls for the perplex analysis of interfacial phenomena. The detailed experimental and theoretical analysis of the coffee-stain formation was undertaken first by Deegan et al. in Refs. 1–2. The authors of Refs 1–2 tested a wide range of experimental conditions and reported formation of ring-like deposits whenever the surface was partially wet by the fluid irrespective of the chemical composition of the substrate. Rings were found in big (~15 cm) and small drops (~ 1 mm) [1–2]. They were found with aqueous and non-aqueous acetone, methanol, toluene, and ethanol solvents [1–2]. Ring-like deposits were found with solutes ranging in size from the molecular sugar and dye molecules to the colloidal (10 µm polystyrene microspheres) and with

---

solute volume fractions ranging from $10^{-6}$ to $10^{-1}$. Likewise, environmental conditions, such as temperature, humidity, and pressure could be extensively varied without affecting the ring. The authors of Refs. 1–2 related the formation of coffee-stain deposits to a couple of main physical reasons, namely: contact line pinning [3–6] and intensive evaporation from the edge of the drop.

However the physical reality is always much more complicated than it is seen through simple physical models. Hu and Larson demonstrated in a series of papers that the thermal Marangoni flow can reduce or completely eliminate the coffee-stain effect for particles dispersed in evaporating octane droplets [7–8]. Indeed, in water droplets Marangoni flows are weak, however in evaporated octane droplets they bring into existence the flow fields, resulting under certain experimental conditions in the inverse coffee-stain effect, namely: the deposit is concentrated at the center of the droplet, and not at its periphery, as predicted by the model reported in Refs. 1–2. This prediction was also verified experimentally in Ref. 9. It was also demonstrated that de-pinning of the triple line results in the inverse coffee-stain effect [10–11].

Later it was shown that flow fields formed under electrowetting also enable controlling formation of "coffee-stain" deposits [12]. The colloidal deposits are also influences by the shape of colloidal particles [13], and it is noteworthy that understanding the "coffee-stain" effect allowed manufacturing ordered colloidal structures [14]. The evolution of the coffee-stain effect in the presence of an insoluble surfactant and non-interacting particles in the bulk was discussed in Ref. 15.

We studied the development of coffee-stain deposits, on oil-infused (impregnated) micro-rough substrates giving rise to a variety of fascinating wetting phenomena and, exerted to the intensive research recently [16–24]. Oil-infused surfaces are featured by the low contact angle hysteresis, low sliding angles and self-healing [25–27]. When droplet of the NaCl water solution was placed on the silicon oil lubricated surface, the contact line was de-pinned and evaporation from the edge of the drop was completely blocked. In contradiction to the physical model developed in Refs. 1–2 the ring deposit was stably observed for a variety of solutions.

## 2. Experimental

Two types of surfaces were coated with a thin layer of polydimethylsiloxane oil (PDMS, supplied by Aldrich) with the molecular mass of 5600–24000 g /mole.

The surfaces were:

1) Glass slides with the dimensions of $80 \times 30$ mm (roughness $0.40 \pm 0.05$ nm, as established with AFM). Glass slides were cleaned with acetone and ethanol and rinsed carefully with distilled water. PDMS oil was spread by the brush, and the slide was kept several minutes in ambient conditions till complete spreading of PDMS.

2) Silicone oil-infused polymer surfaces that have been manufactured as follows. Polypropylene substrates with a thickness of 25 µm were coated with polycarbonate (PC) film under the fast dip-coating process according to the protocol described in detail in [28–29]. As a result, PC film with typical honeycomb porous self-assembly pattern was obtained. The average radius of pores was about 1.5 µm (for the SEM images of porous structures see Refs 28–29). The average depth of pores as established by AFM was about 1 µm. The use of honeycomb surfaces facilitated manufacturing stable silicone oil-infused surfaces.

The dynamic viscosity of PDMS was measured using the Ostwald type viscometer in thermostatic bath at $25°C$. It was established as $\eta = (2.62\pm0.05)\,10^{-3}\,\text{Pa}\times\text{s}$. The thickness of PDMS oil layer was established by weighing for both types of solid substrates as $50\pm2$ µm.

Droplets of aqueous solutions of NaCl, KCl, $Na_2SO_4$ and $CaCl_2$ were placed on both types of substrates (see **Figure 1**). Bi-distilled water (the specific resistivity $\rho \cong 18\cdot\text{M}\Omega\times\text{cm}$) was used for preparing solutions. The initial molar concentrations of solutions were in the range of 1–2.5 M. 5–300 µl droplets of the solution were placed on the PDMS oil and evaporated at room temperature. Images were taken through the process every 5 minutes with digital microscope BW1008-500x. The experiments were carried at ambient conditions ($t = 21^0 C$; relative humidity $RH = 35-40\%$). The experiments were repeated 10 times. Thermal imaging of the process was carried out with the Therm-App TAS19AQ-1000-HZ thermal camera equipped with a LWIR 6.8 mm f/1.4 lens.

3. **Results and discussion**

In all experiments, we observed formation of the typical "coffee-stain" precipitate at the circumference of evaporated droplets as depicted in **Figure 2**. (in the vicinity of the contact line, which is strictly speaking is not "a triple line", as shown in **Figure 3**). The complete repeatability of the observation of the ring-like precipitate is noteworthy. It should be emphasized that the coffee-stain-like precipitate was formed on both types of the studied solid supports, namely porous and flat, mentioned in the Experimental Section. The characteristic time of evaporation was 30–150 min, depending on the initial volume of a droplet. When a water droplet was placed on the silicone oil, it was coated with the silicone oil, as depicted schematically in **Figure 3**. This wetting situation is well explained by the analysis of the spreading parameter *S* governing the wetting situation [30–32]:

$$S = \gamma - (\gamma_{\text{oil}} + \gamma_{\text{oil/water}}), \qquad (1)$$

where $\gamma \cong 70 \text{ mJ/m}^2$, $\gamma_{\text{oil}} \cong 20 \text{ mJ/m}^2$, $\gamma_{\text{oil/water}} \cong 23-24 \text{ mJ/m}^2$ are the interfacial tensions at water/vapor, oil/vapor and oil/water interfaces respectively (interfacial tensions are taken from the literature data [33]). Substituting the aforementioned values in Exp. 1, we obtain $S > 0$; in this case, the silicone oil is expected to coat the water droplet, at least partially [23]. It should be stressed, that PDMS did not coat a water droplet completely, as shown in **Figures 1, 3**. A hole with the diameter of *ca* 1-4 mm (depending on the size of the drop and the measured oil) remained uncoated, enabling relatively slow evaporation of solutions from the top of a droplet (the characteristic time of complete evaporation of 5–50 µl aqueous solutions droplets deposited on solid substrates was of the order of magnitude of 20–90 min).

Now address the formation of the coffee-stain precipitate which is paradoxical and contra-intuitive for two-reasons: there is no pinning of the contact line on the oil-infused substrates and the evaporation of a droplet is completely blocked at its edge by silicon oil. Thus, main physical conditions, responsible for the formation of the coffee-stain, namely pinning of the contact line and intensive evaporation from the edge of the drop are violated [1–2]. The reasonable question is: what is physical mechanism, constituting the ring-like deposits, observed on oil-impregnated surfaces, depicted in **Figure 2**?

We propose following mechanism involving co-occurrence of thermal and solutal Marangoni flows [34–35], driven by evaporation of a droplet and resulting in temperature and concentration gradients. When evaporation occurs from the top of a droplet the axisymmetric distribution of temperatures at the surface of a droplet is expected, with temperatures decreasing from the axis to the circumference of a droplet. Indeed, such a temperature profile was registered, as shown in **Figure 4**. The maximal temperature change across the droplet as high as 2.5° was observed. This temperature jump promoted the Marangoni temperature flow proceeding from the perimeter to the top of droplet, as shown in **Figure 3**.

In addition evaporation creates the area enriched with a solute which is close to the top of a droplet, as depicted in **Figure 3**. Thus, the soluto-capillary flow starts directed towards the top of a droplet [34–35]. Consider that both thermo- and soluto-capillary flows drive water to the top of a droplet, namely the only area, where the evaporation takes place. Thus, areas enriched by the solute, which are close to the perimeter of a droplet, are created (see **Figure 3**), giving rise eventually to the formation of the ring-shaped, coffee-stain precipitate.

The question is: what kind of Marangoni flow (thermal or solutal) is responsible for the formation of the coffee-stain precipitate? In order to answer this question, consider that the thermal Marangoni flow is characterized with the thermal Marangoni number:

$$Ma_T = \frac{\left|\frac{\partial \gamma}{\partial T}\right|\Delta T d}{\eta \kappa} \quad . \tag{2}$$

Whereas, the soluto-capillary flow is quantified by the soluto-capillary Marangoni number, supplied by:

$$Ma_C = \frac{\left|\frac{\partial \gamma}{\partial c}\right|\Delta c d}{\eta D} \quad , \tag{3}$$

where $\kappa$ and $\eta$ are the thermal diffusivity and viscosity of the solution correspondingly, $D$ is the diffusion coefficient of a salt in water at ambient conditions, $d$ – is the characteristic length, which is in our case 2–10 mm, depending on the volume of a droplet [34–35]. The interrelation between thermo- and soluto-capillary flows is described by the dimensionless number:

$$\zeta = \frac{Ma_T}{Ma_C} = \frac{\left|\frac{\partial \gamma}{\partial T}\right|\Delta T D}{\left|\frac{\partial \gamma}{\partial c}\right|\Delta c \kappa} \tag{4}$$

Substituting :
$\left|\frac{\partial \gamma}{\partial T}\right| = 0.167 \frac{N}{m \times K}; \left|\frac{\partial \gamma}{\partial c}\right| = 3.16 \frac{N \times kg}{m \times mol}; \Delta T = 1.5 K; \Delta c = 4 \frac{mol}{kg}; D = 1.47 \times 10^{-9} \frac{m^2}{s}; \kappa = 14 \times 10^{-8} \frac{m^2}{s}$

into Eq. 4 yields the estimation: $\zeta \cong 2 \times 10^{-4}$ (for the numerical values of the parameters appearing in Eq. 4 see Refs. 36–40). The estimation brings to the conclusion that the formation of the coffee-stain effect is mainly governed by the soluto-capillary Marangoni flow. It is noteworthy that the value of the dimensionless number $\zeta = \frac{Ma_T}{Ma_C}$ is independent on the characteristic length $d$; thus the effect of the formation of the coffee-stain ring will be driven by the soluto-capillary Marangoni flow for any reasonable volume of a droplet. Indeed, the coffee-stain precipitates were observed for 5–300 µl droplets, in other words, the ring-like deposit was formed for both of $R \geq l_{ca}$ and $R \leq l_{ca}$ experimental situations, where $R$ is the characteristic dimension of the droplet, and $l_{ca}$ is the capillary length.[30-32]

It should be mentioned that the thermo-capillary Marangoni flows, negligible for solutions, may be essential for suspensions, i.e. droplets containing colloidal particles [1-2], to be studied on our future work. The development of the quantitative model of the process class also calls for future investigations [41].

## 4. Conclusions

Oil-impregnated porous films attracted considerable attention of investigators in a view of their applications as multifunctional, antifouling, self-healing surfaces [17-20]. Water droplets contacting oil-impregnated surfaces inevitably contain salts, which may give rise to the coffee-stain effect [1-2, 10-11], deteriorating surfaces and consequently destroying their useful properties [41]. Our investigation is devoted to the study of the coffee-stain effect on oil-impregnated surfaces. The formation of coffee-stain ring-like deposits has been observed when 5–300 μl droplets of aqueous solutions of NaCl, KCl, $Na_2SO_4$ and $CaCl_2$ were placed on the silicone oil-impregnated solid surfaces. It is generally accepted that the formation of coffee-stain precipitates is due to the coupling of contact line pinning and intensive evaporation from the edge of the drop [1–2, 6, 10–11]. However in the case when droplets are placed on oil-infused substrates both of these mechanism do not work, namely: the contact line is de-pinned and evaporation from the edge of a droplet is blocked by the silicone oil, coating the periphery of a droplet. In spite of this, the formation of the coffee-stain deposits was observed for a broad range of molar concentrations of various salts and within a broad range of volumes of droplets. We related the formation of ring-like deposits to the co-occurrence of soluto-capillary and thermo-capillary Marangoni flows [34–35]; with the decisive role of the soluto-capillary Marangoni flow. The dimensionless number $\zeta = \dfrac{Ma_T}{Ma_C}$, introduced in the paper, describing interrelation between soluto-capillary and thermo-capillary Marangoni flows, is independent on the characteristic dimension of a droplet; hence, the effect of the formation of the coffee-stain ring will be driven by the soluto-capillary Marangoni flow for various volumes of droplets.

**Acknowledgements**
The authors are indebted to Mrs. Yelena Bormashenko and Dr. Albina Musin for their kind help in preparing this manuscript.
Acknowledgement is made to the donors of the Israel Ministry of Absorption for the partial support of the scientific activity of Dr. Mark Frenkel.

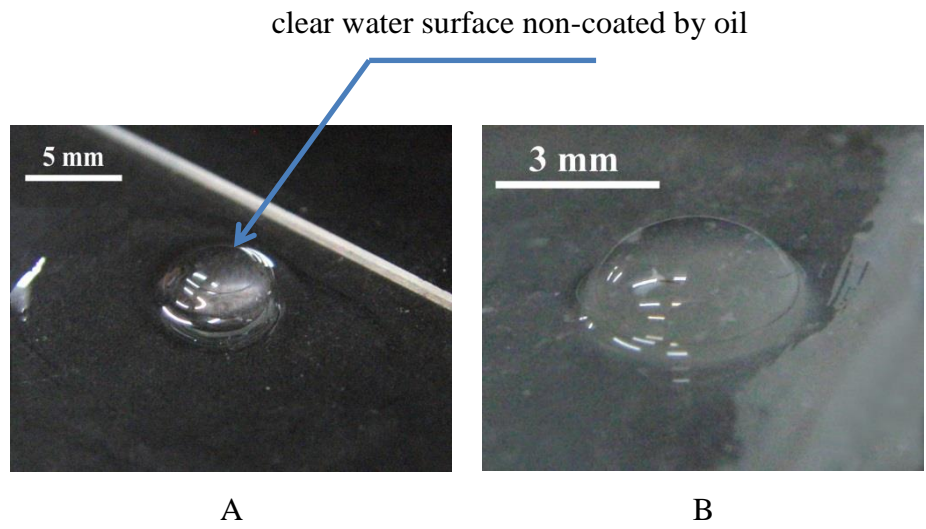

**Figure 1**. A. 30 µl water droplet placed on the silicone oil spread on the glass slide surface. B. 30 µl water droplet placed on the silicone oil-infused porous polymer surface.

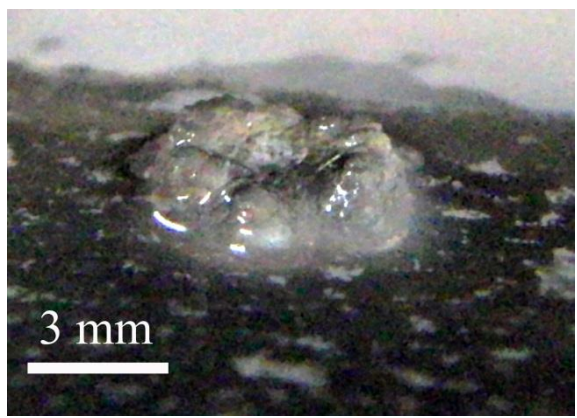
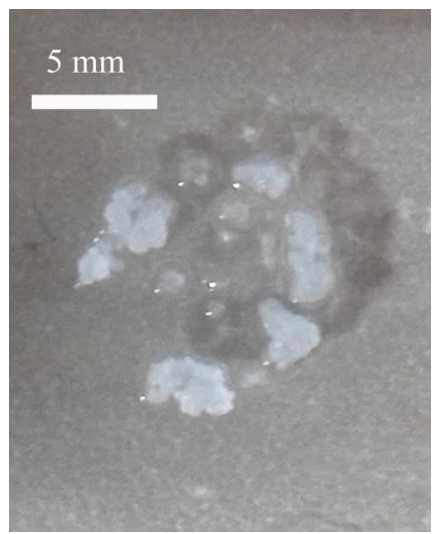

A                                                       B

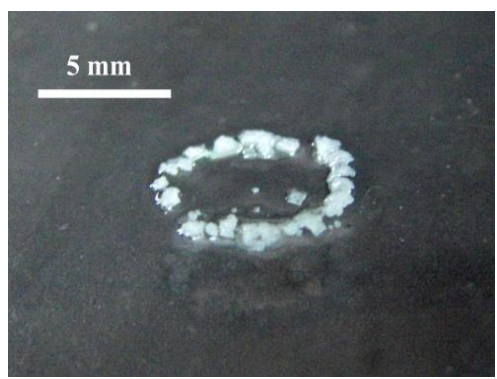

C

**Figure 2**. Coffee-stain deposits, observed on oil-impregnated surfaces. A. The salt ring-like deposit formed after evaporation of 20μl droplet of 2M Na$_2$SO$_4$ solution, placed on the PDMS impregnated surface (on PC surface). B. Crystals of salt observed after evaporation of droplet with volume of 40μl of 2.5M NaCl (on glass). C. Crystals of salt after evaporation of droplet with volume of 50μl of 2.5M NaCl (on glass).

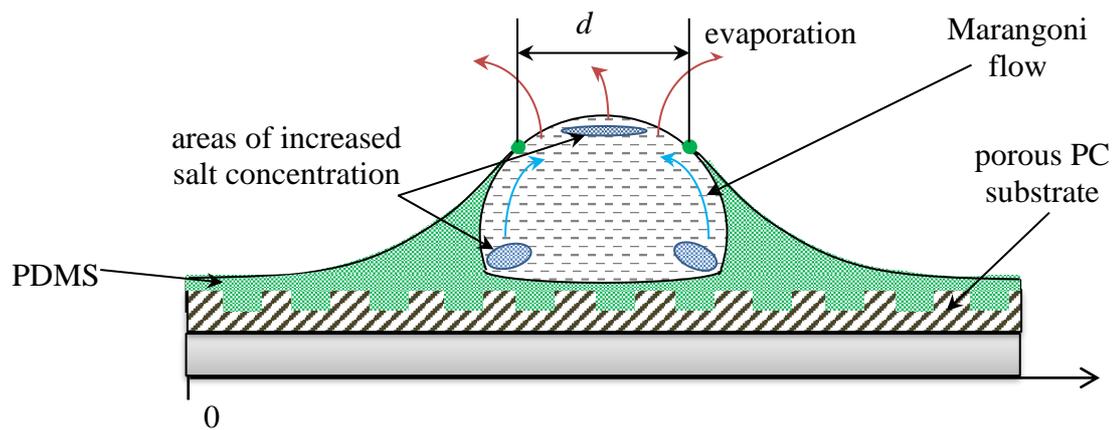

**Figure 3**. Scheme of formation of the coffee-stain deposit under the evaporation of droplets, containing aqueous solutions of salts, placed on the oil-impregnated surfaces.

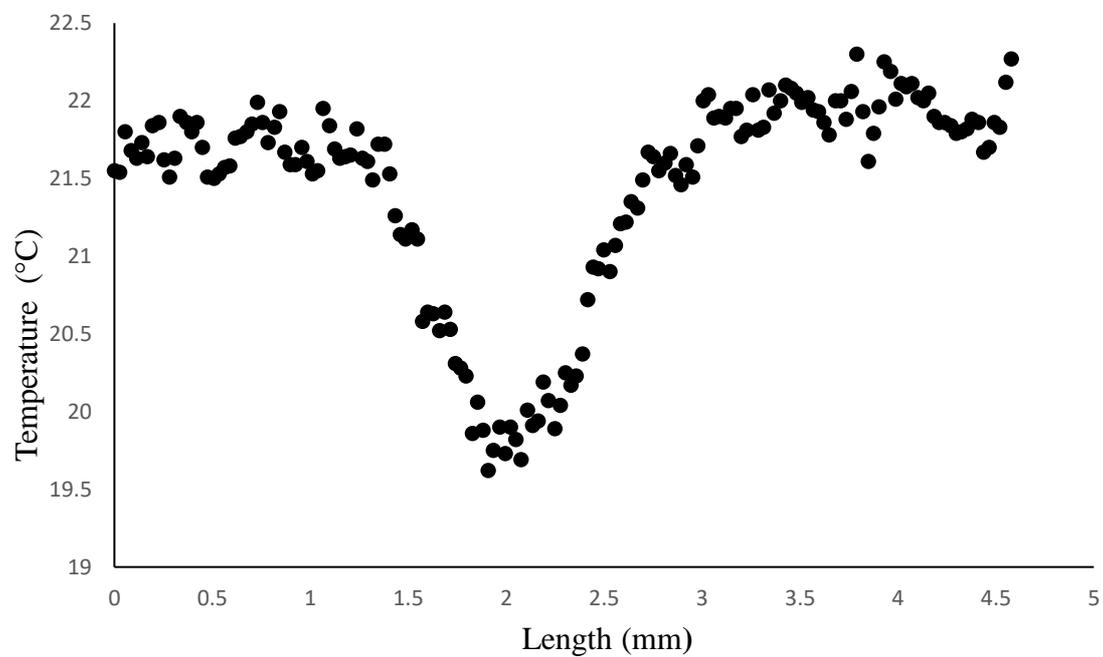

**Figure 4**. The lateral temperature profile of the droplet with volume of 10 μl of 2.5 M NaCl solution. Zero point of the abscissa is shown in **Figure 3**.